# ICT's role in e-Governance in India and Malaysia: A Review

[1]Ganesh Ch Deka, [2]Jasni Mohamad Zain, [3]Prabhat Mahanti

[1]Assistant Director of Training, in the Directorate General of Employment and Training, Ministry of Labour & Employment, New Delhi, India
[1]ganeshdeka2000@gmail.com

[2]Dean of Faculty of Computer Systems & Software Engineering, University Malaysia
[2]jasni@ump.edu.my

[3]Professor of Computer Science at the University of New Brunswick, Canada
[3]pmahanti@gmail.com

## Abstract

*Information and Communication Technologies (ICTs) play a key role in Development & Economic growth of the Developing countries of the World. Political, Cultural, Socio-economic Developmental & Behavioral decisions today rests on the ability to access, gather, analyze and utilize Information and Knowledge. Government of India is having an ambitious objective of transforming the citizen-government interaction at all levels to by the electronic mode by 2020.Similarly according to the Vision 2020-The Way Forward presented by His Excellency YAB Dato' Seri Dr Mahathir Mohamad at the Malaysian Business Council "By the year 2020, Malaysia can be a united nation, with a confident Malaysian society, infused by strong moral and ethical values, living in a society that is democratic, liberal and tolerant, caring, economically just and equitable, progressive and prosperous, and in full possession of an economy that is competitive, dynamic, robust and resilient". This paper presents a comparative study and review relating to e-Governance and application of ICT development between India & Malaysia.*

## 1. Introduction

Governance refers to the process of interaction among the Government, Business, and Civil society to manage their Political, Social and Economic environment [10].
There are six dimensions of Governance:

i. Voice & Accountability
ii. Political Stability and Lack of Violence/Terrorism
iii. Government Effectiveness
iv. Regulatory Quality
v. Rule of Law, and
vi. Control of Corruption

Higher values on these dimensions thus indicate better governance ratings. The developed countries of the World are having much higher ratings than the developing countries.ICT has showed its revolutionary power as a key catalyst for change, modernization and innovation and this trend will only accelerate going forward. ICT's use to induce changes in governance presents a big potential in opening up governance processes, but it should be preceded by good governance foundations. Traditional governance processes, which have already been impeded by lack of citizen's participation, poor performance of government services, lack of accountability and transparency, have to be revitalized in conjunction with the introduction of electronic governance mechanisms. The legislature, judiciary and administration may apply e-governance, in order to improve internal efficiency, the delivery of public services, or processes of democratic governance. It also refers to the citizen to government interface including the feedback of policies. At the same time, citizens have to undergo a mind-set change, overcome the "culture of fear" or the "culture of non- confrontation" skepticism and lack of communal involvement in seeing to the welfare of their society [5].

For speeding up the e-Governance implementation across the various arms of Government at National, State, and Local levels due cognizance has been taken to adopted common vision and strategy having the potential of huge costs savings in presenting a seamless view of Government to citizens by;

i. Sharing of core and support infrastructure
ii. Enabling interoperability through standards

E-governance has not only the ability to handle momentum and complexity but also to underpin the regulatory reform. Even though ICT is not substitute for good policy, it empowers the citizens to question the actions of regulators and brings systemic issues to the forefront.

In 1991, the Malaysia Prime Minister enunciated Malaysia's desire to become a developed country by 2020 i.e. by the end of its 11th Development Plan.

The use of ICT the public sector is increasingly being adopted in Malaysia. Malaysia is ranked 10th for ICT readiness of the society as a whole. ICT plays a critical role in its Wawasan 2020 (Vision 2020) plan for Malaysia to become a high-income economy by 2020[3].

The Multimedia Super Corridor (MSC Malaysia) was established in 1996 with the aspiration of becoming a global hub for ICT and multimedia innovation, operation and services and to transform Malaysia into a knowledge-economy and achieve developed nation status in line with Vision 2020[4].

Government of India is having an ambitious objective of transforming the citizen-government interaction at all levels to by the electronic mode (e-Governance) by 2020.

The rest of this paper is organized into 7(seven) Sections. Section-2 is about e-Government, in Section-3 we have compared Malaysia and India based on the data from the "The Global Technology Report 2010-2011", Section-4 is about the ICT Laws in India & Malaysia. Section-5 is about ongoing Government ICT initiatives in Malaysia & India. Section-6 is about our findings & observations and finally Section-7 discusses the future scope of study and research direction in this area.

## 2. e-Governance in India

The first e-Governance initiative in India was the Computerization of Government Departments. Present e-Governance initiatives will be encapsulating the finer points of Governance such as Citizen Centricity, Service Orientation & Transparency.

Government of India approved the National e-Governance Plan (NeGP) [3] on 18$^{th}$ May 2006 comprising of the following:

i. Vision
ii. Approach
iii. Strategy
iv. Key components
v. Implementation methodology &
vi. Management Structure

Different accessible or continuing projects in the Mission Mode Project (MMP) [2] category under the Central Ministries, States & State Departments would be properly improved and enhanced to align with the objectives of NeGP.

The followings are the various MMP under the NeGP:
1) Central MMPs
   i. Banking
   ii. Central Excise & Customs
   iii. Income Tax (IT)
   iv. Insurance
   v. MCA21
   vi. National Citizen Database
   vii. Passport
   viii. Immigration, Visa and Foreigners Registration & Tracking
   ix. Pension
   x. e-Office
2) State MMPs
   i. Agriculture
   ii. Commercial Taxes
   iii. e−District
   iv. Employment Exchange
   v. Land Records
   vi. Municipalities

    vii. Gram Panchayats
    viii. Police
    ix. Road Transport
    x. Treasuries
3) Integrated
    i. CSC
    ii. e-Biz
    iii. e-Courts
    iv. e-Procurement
    v. EDI For e-Trade
    vi. National e-governance Service Delivery Gateway
    vii. India portal

### 3. ICT in Malaysia

The e-Government initiative was launched to lead the country into the Information Age, to improve how the government operates internally as well as how it delivers services to the people of Malaysia [17].

It seeks to improve the convenience, accessibility and quality of interactions with citizens and businesses. At the same time, it will improve information flows and processes within government to improve the speed and quality of policy development, coordination and enforcement.

To accelerate the objectives of Vision 2020, a path has been defined through 7(seven) innovative Flagship Applications. These applications are engineered to start the MSC initiative and create a multimedia paradise for innovative producers and users of multimedia technology. Both local and foreign companies work with various government agencies to enhance the socio-economic development of Malaysia.

Flagship Applications of Multimedia Super Corridor of Malaysia are:
    i. Electronic Government
    ii. Multipurpose Card
    iii. Smart School
    iv. Tele-health
    v. R&D Clusters
    vi. E-Business
    vii. Technopreneur Development

Under the e-Government flagship, seven main projects were identified to be the core of the e-government applications. The e-Government projects are [13]:
    i. Electronic Procurement (eP)[15]
    ii. Project Monitoring System (PMS)
    iii. Electronic Services Delivery(eServices)
    iv. Human Resource Management Information System (HRMIS)
    v. Generic Office Environment (GOE)
    vi. E-Syariah and
    vii. Electronic Labour Exchange (ELX).

Besides these seven main projects under e-government flagships, several government agencies has taken initiatives to introduced online services for the public projects, aims to increase the ease and efficiency of public service to the people.

Among others were Public Services Portal (my Government), e-Tanah, e-Consent, e-Filing, e-Local Government (e-PBT), e-Kehakiman, Custom Information System (SMK), Pensions Online Workflow Environment (POWER), and Training Information System (e-SILA).

Four strategic thrusts have been identified in order for the government to realize its vision for 2020. That includes [14]:
- "1 Malaysia, People First, Performance Now"
- Government Transformation Programme (GTP)
- Economic Transformation Programme (ETP)
- 10$^{th}$ Malaysia Plan

Various programmes are developed to address the ICT requirements for the government in the above-mentioned areas over the next five years.

### 4. The Global Technology Report 2010-2011

The Global Information Technology Report [2] series tried to cast light on the evolution of ICT over the last decade as well as raising awareness about the importance of ICT diffusion and leveraging for increased development, growth, and better living conditions. The methodological framework of the Networked Readiness Index (NRI) has mapped out the enabling factors driving networked readiness, which is the capacity of countries to fully benefit from new technologies in their competitiveness strategies and their citizens' daily lives. The Index has allowed private and public stakeholders to monitor progress for an ever-increasing number of economies all over the globe as well as to identify competitive strengths and weaknesses in national networked readiness landscapes.

The following are the comparative data between Malaysia & India in terms of some of the vital parameters in terms of ICT application in Development.

A.     The Networked Readiness Index

Networked Readiness Index (NRI) is a comprehensive assessment of the present state of networked readiness of the Country.

These relate to:

i. The emerging Internet economy
ii. Communities to be built around digital highways
iii. The promise of technology
iv. ICT's growing impact on poverty reduction
v. ICT's contribution to meeting the decade's challenges
vi. Localization 2.0
vii. ICT for an effective social strategy
viii. The creation of a fiber future and its regulatory challenge, and
ix. Mobile banking in the emerging world

The following table shows the ranking of India & Malaysia in NRI:

| Country | Rank in the Income group | Score | *Income group | Rank in the Income group |
|---------|--------------------------|-------|---------------|--------------------------|
| Malaysia | 28 | 4.74 | Upper-Middle | 1 |
| India | 48 | 4.03 | Lower Middle | 3 |

*Country classification by Income group is from the World Bank (situation as on December 2010)

The networked readiness for Asia and the Pacific is by and large positive. Malaysia is the only upper-middle-income country within the top 30 overall. The success stories are a source of inspiration for a number of underperformers in the region, including Timor-Leste (136th), Nepal (131st), Bangladesh (115th) and Pakistan (88th).The data shown above reveals that in terms of Network Readiness Malaysia is in better position than India. India's placement is dragged down by its poor and more generally by the poor marks in most education-related variables included quality of its soft and hard infrastructures (81st).

B.     Usage Sub-index

The last component of the NRI measures the actual ICT usage by an economy's main social actors and includes a total of 20 variables. This sub-index will progressively evolve toward capturing ICT impact in terms of inclusive society, business innovation, and better governance.

| Country | Score (Out of 7) | Individual Usage | | Business Usage | | Government Usages | |
|---------|------------------|------|-------|------|-------|------|-------|
| | | Rank | Score | Rank | Score | Rank | Score |
| Malaysia | 4.53 | 45 | 4.26 | 15 | 4.24 | 11 | 5.10 |
| India | 3.34 | 98 | 2.83 | 45 | 3.38 | 47 | 3.82 |

Other parameters having impact of ICT to the society are compared in the table below (2009–2010 weighted average.) [1=nonexistent; 7=well developed]

| Parameter | Country | Rank | Score(out of 7) | Mean |
|---|---|---|---|---|
| Importance of ICT to Government vision to future | Malaysia | 11 | 5.13 | 4.01 |
| | India | 32 | 4.61 | |
| Government prioritization of ICT | Malaysia | 12 | 5.76 | 4.66 |
| | India | 35 | 5.31 | |
| Laws relating to ICT | Malaysia | 26 | 5.08 | 3.98 |
| | India | 39 | 4.55 | |
| Impact of ICT on new organizational models | Malaysia | 13 | 5.25 | 4.15 |
| | India | 34 | 4.73 | |
| Impact of ICT on new services & products | Malaysia | 23 | 5.36 | 4.50 |
| | India | 36 | 5.08 | |
| Government success in ICT promotion | Malaysia | 14 | 5.30 | 4.34 |
| | India | 22 | 5.17 | |
| ICT use and Government efficiency | Malaysia | 15 | 5.37 | 4.30 |
| | India | 41 | 4.69 | |
| Government Online service Index | Malaysia | 16 | 0.63 (out of 1) | |
| | India | 53 | 0.37 | |
| Internet access in schools | Malaysia | 36 | 4.96 (out of 7) | 4.06 |
| | India | 70 | 3.83 | |
| Impact of ICT on access to basic services | Malaysia | 15 | 5.50 | 4.48 |
| | India | 42 | 4.88 | |
| Households with a personal computer | Malaysia | 58 | 38.7% | |
| | India | 118 | 4.4% | |

## 5. ICT Acts dealing with the e-Governance issues in India and Malaysia

A. The followings are the ICT acts prevailing in India.

| Name of the Act | Purpose |
|---|---|
| THE INFORMATION TECHNOLOGY ACT, 2000. | • To make the use of ICT effective the parliament of India has passed this Act. In this act the CHAPTER III represents the-ELECTRONIC GOVERNANCE where Section-8 indicates the Publication of rule, regulation etc. in Electronic Gazette.<br>• Any Rule, regulation, order, bye-law, notification or any other matter published in the Official Gazette should also be published in the Electronic Gazette.<br><br>CHAPTER X represents THE CYBER REGULATIONS APPELLATE TRIBUNAL where Section-48 indicates the Establishment of Cyber Appellate Tribunal.<br>• The Central Government shall, by notification, establish one or more appellate tribunals to be known as the Cyber Regulations Appellate Tribunal.<br>• The Central Government shall also specify, in the notification referred to in subsection<br>• The matters and places in relation to which the Cyber Appellate Tribunal may exercise jurisdiction<br>The Office and the Court Room of the Cyber Regulation Appellate Tribunal in New Delhi for handling the liability to check Cyber fraud, Cyber crime & Cyber terrorism was inaugurated on 27th July 2009, by the then Chief Justice of India Mr. Justice K.G. Balakrishnan. |

| | | |
|---|---|---|
| THE RIGHT TO INFORMATION ACT 2005 | | In this act the CHAPTER II (Sec-2 & Sec-4) represents the Right to information and obligations of public authorities<br><br>Section-2: It shall be a constant endeavour of every public authority to take steps in accordance with the requirements of clause (b) of sub-section (1) to provide as much information suo motu to the public at regular intervals through various means of communications, including Internet, so that the public have minimum resort to the use of this Act to obtain information.<br>Section-4: All materials shall be disseminated taking into consideration the cost effectiveness, local language and the most effective method of communication in that local area and the information should be easily accessible, to the extent possible in electronic format with the Central Public Information Officer or State Public Information Officer, as the case may be, available free or at such cost of the medium or the print cost price as may be prescribed.<br>Explanation- For the purposes of Sub-Sections-4 "disseminated" means making known or communicated the information to the public through notice boards, newspapers, public announcements, media broadcasts, the Internet or any other means, including inspection of offices of any public authority. |
| The Information Technology (Amendment) Act 2008, Section 69 | | This act authorize the Central Government/State Government/its authorized agency to intercept, monitor or decrypt any information generated, transmitted, received or stored in any computer resource if it is necessary or expedient so to do in the interest of the sovereignty or integrity of India, defense of India, security of the State, friendly relations with foreign States or public order or for preventing incitement to the commission of any cognizable offence or for investigation of any offence. |

B.  ICT Acts in Malaysia[4]

| Sl No | Name of the Act | Purpose |
|---|---|---|
| 1 | Act 563: COMPUTER CRIMES ACT 1997 [REPRINT 2002] | a) Seeks to make it an offence for any person to cause any computer to perform any function with intent to secure unauthorized access to any computer material.<br><br>b) Seeks to make it a further offence if any person who commits an offence referred to in item (a) with intent to commit fraud, dishonesty or to cause injury as defined in the Penal Code.<br><br>c) Seeks to make it an offence for any person to cause unauthorized modifications of the contents of any computer.<br><br>d) Seeks to provide for the offence and punishment for wrongful communication of a number, code, password or other means of access to a computer.<br><br>e) Seeks to provide for offences and punishment for abetments and attempts in the commission of offences referred to in items (a), (b), (c) and (d) above.<br><br>f) Seeks to create a statutory presumption that any person having custody or control of any program, data or other information when he is not authorized to have it will be deemed to have obtained unauthorized access unless it is proven otherwise. |
| 2 | ELECTRONIC GOVERNMENT ACTIVITIES BILL 2007 | The main motto of this Act is to provide the legal recognition of electronic messages in dealings between:<br><br>- The Government and the Public<br>- The use of the electronic messages to fulfill legal requirements and to enable and facilitate the dealings through the use of electronic means and other matters connected therewith. |

| 3 | Communications & Multimedia Act 1998 | The objects of this Act are:<br>• To promote national policy objectives for the communications and multimedia industry;<br>• To establish a licensing and regulatory framework in support of national policy objectives for the communications and multimedia industry;<br>• To establish the powers and functions for the Malaysian Communications and Multimedia Commission; and<br>• To establish powers and procedures for the administration of this Act. |
|---|---|---|
| 4 | Malaysian Communication & Multimedia Commission Act 1998 | An Act to provide for the establishment of the Malaysian Communications and Multimedia Commission with powers to supervise and regulate the communications and multimedia activities in Malaysia, and to enforce the communications and multimedia laws of Malaysia, and for related matters. |
| 5 | Digital Signature Act 1997 | An Act to make provision for, and to regulate the use of, digital signatures and to provide for matters connected therewith |
| 6 | Telemedicine Act 1997 | An Act to provide for the regulation and control of the practice of telemedicine; and for matters connected therewith. |
| 7 | Optical Discs Act 2000 | An Act to provide for the licensing and regulation of the manufacture of optical discs, and for matters connected therewith. |

## 6. Ongoing e-Governance Initiatives in Malaysia & India

The followings are the ongoing e-Governance initiatives in Malaysia & India:

| Malaysia | India |
|---|---|
| 1. Malaysian Communications and Multimedia Commission (MCMC) or Suruhanjaya Komunikasi dan Multimedia Malaysia (SKMM): The vision of MCMC of SKMM is to be "A globally competitive, efficient and increasingly self-regulating communications and multimedia industry generating growth to meet the economic and social needs of Malaysia [7]. | 1. National e-Governance Plan (NeGP) consists of 27 Mission Mode Projects (MMPs) of State Governments & Central Government. There are 8 Integrated Mission Mode Projects (MMPs) which are to be implemented by the State Governments & Central Government combined |
| 2. Government of Malaysia is planning to connect 300 villages in Pahang, the third largest Malaysian state with 1,435,701 inhabitants by the end of 2011 with Wi-Fi. Datuk Seri Dr Rais Yatim, Information Communication and Culture Minister said the Kampung (village) Wi-Fi programme by the Communications and Multimedia Commission (MCMC) involved the construction of 863 telecommunication towers in stages. The project would cost an estimated RM25, 000 (US$8225) to RM30, 000 (US$9870) each depending on its distance from the town centre or community centre equipped with Internet wireless facility. In total 1,000 Kampung WiFi would be launched nationwide by the year end. | 3. There are 13 Data Centers are up and 26 State Wide Area Network (SWAN) are operational. However the States without have their own infrastructure can take the benefit of Cloud Computing like the state of Jammu & Kashmir.<br>4. The mobile subscribers in India are likely to reach 1.159 billion by the end of 2013 [11]. Considering the rapid penetration rate of mobile phones in rural area the integration of Mobile Services Delivery Gateway and UIDAI [Aadhaar] will be foundation of the m-Government for the inclusive Development of India. |

## 7. Conclusion

There has been a paradigm shift emerging over the past few years, how governments work and use the Web and ICT to deliver better services to their constituencies. The new approach is known as Open Government, means rethinking how to govern and how the administrations should adapt their procedures to meet the demands and necessities of the citizens.

The costs associated with telecommunication infrastructure and human capital continues to hold back e-governance development. However, effective strategies and legal frameworks can recompense significantly, even in least developed countries. There is an urgent need to integrate ICT training from the earliest educational level of the community, so that the technology becomes familiar as they grow along specifically in India.

Malaysia's unprecedented growth and dramatic reduction in poverty should make all developing countries take notice of their systems. Malaysia is the only developing country that declared that it wants to be developed country and is on track to becoming one. It has done this with long spells of surplus budgets and a good record of plan implementation. In fact, it managed to have a soft landing during the East Asian Crisis because of the flexibility and breathing space provided by these prudent policies and sound public management systems [9].The comparative study revealed that Malaysia is in a better position than India in many aspects so far as the overall implementation of ICT and e-Governance is concerned. India needs focused ICT Acts & policies like Malaysia.

Implementation of IT Act 2000 is yet in a nebulous stage though the Act was enacted 10 years back. The biggest challenge of deploying e-Governance is not technology, but ability of various stakeholders to work together to manage the change introduced by ICT.

There is an urgent need of ICT Act to address all the Legal issues pertaining to ICT.

However, most of the Government organizations in India either not developed their Website or the few which have developed their Websites were not updated regularly as a result the requisite information regarding the activities of the organizations is not readily available to the cliental group online.

If the websites of the Government Departments/ Ministries were updated regularly it will not only enhance the transparency but will be saving the valuable time otherwise spent entertaining the enquiries of the public as well replying the questions asked under the Right to Information Act 2005.

Electronic form of communications is rarely used for Official communications in India as of now [8].

The 12 digit UID [Aadhaar] numbers issued by Unique Identification Authority of India (UIAI) can be used as single password to access all e-Government services like Sing-Pass (Singapore Personal Access) in Singapore for e-Citizen, the one-stop online portal for all government services and information.

For effective use of mobile communication, a common mobile public service framework incorporating the 5 principles like Interoperability, Security, Openness, flexibility and scalability [12] is necessary. There are lots of scopes for Research & Development in these fields considering the scope for implementing the m-Government in India. Most of mobile users in India are first time users' who neither used Laptop or Desktop. Non-availability of ICT contents in local language is an added handicap.

## 8. Future ahead

There is great scope for study and research in the field of e-Governance, Cyber Law & Implementation of Information Technology Acts. The followings are the future ICT vision ahead in India & Malaysia:

| Malaysia | India |
|---|---|
| According to the "The Way Forward-Vision 2010" Malaysia will be a "Fully Developed Country" joining the group of 19 countries that are generally regarded as "Developed countries". The projected GDP of Malaysia will be around 920 billion Ringgit. To achieve this goal they must work efficiently, effectively, with fairness and dedication. Government of Malaysia is planning to connect 300 villages in Pahang, the third largest Malaysian state by the end of 2011 with Wi-Fi. The project would cost an estimated RM25, 000 (US$8225) to RM 30, 000 (US$9870) -each depending on its distance from the town centre or community centre equipped with Internet wireless facility. | In India the m-Government cell is being created under the Department of Information Technology's(DITs) e-Governance unit with a mix group of 1,500 officials, comprising members from the Academy, Industry stake holder, civil society(Non–Government organizations) and officials from DIT and Department of Telecommunication. The aim is to deliver all services presently available through computers through Mobile Technology. |
| The urban broadband penetration in Malaysia is currently 57.4%. Government of Malaysia is planning to make it 60% by the year end of 2011. | India's Telecom Commission proposal to create a US$4.5 billion National Optical Fibre Network (NOFN) has been approved by the Department of Telecom (DoT), announced Shri Kapil Sibal, Minister of Communication & Information Technology.<br>This NOFN will extend the country's existing fibre optic network from the district level to the village level, giving the country of 1.2 billion people services like e-education, e-health, e-banking and also reduce migration of rural population to urban areas[16]. |
| | Department of Information Technology (DIT), Government of India, to unveil G-Cloud (Government Cloud Computing) policy by 2012.<br>For the effective implementation of NeGP the Cloud computing will play a big role. The community cloud may be used to deliver plethora of e-Government services at the doorstep of the rural community.<br>As part of 100 day agenda, Ministry of Information Technology, Government of India is planning to come up with Electronic Service to be passed in parliament as an electronic service delivery Act (ESDA) by the end of this year. This bill intends to make it mandatory for all government departments and ministries to deliver public services only in electronic mode from a cut date and that cut date will be fixed by the concerned department or the ministry depending upon their level of readiness. |